\documentstyle[aps,prl]{revtex}
\def\slash#1{\setbox0=\hbox{$#1$}#1\hskip-\wd0\hbox to\wd0{\hss\sl/\/\hss}}
\begin{document}
\centerline{\large\bf Dynamical Gauge Symmetry Breaking in ${SU(3)_L\otimes U(1)_X}$ }

\centerline{\large\bf Extension of the Standard Model 
}

\bigskip

\begin{center}
{\large\bf Prasanta Das  and Pankaj Jain 
}\\

\bigskip
Physics Department, I.I.T., Kanpur 208016, India\\

pdas@iitk.ac.in, pkjain@iitk.ac.in
\end{center}

\begin{abstract}
We study the SU(3)$_L\otimes$ U(1)$_X$ extension of the standard 
model with a strong U(1) coupling. 
We argue that current experiments
limit this coupling to be relatively large. The model
is dynamically broken to the Standard SU(2)$_L\otimes $U(1) model
at the scale of a few TeV with all the extra gauge bosons and the 
exotic quarks acquiring masses much larger than the scale of
electroweak symmetry breaking. Furthermore we find that the model 
naturally displays 
the top  condensation mechanism for the dynamical breakdown of 
electroweak gauge symmetry.  
By adding a $SU(3)_L$ singlet quark $\chi$, we find that the model predicts
the correct mass for the top quark and the $W,Z$ bosons.  Based of the
dynamical symmetry breaking mechanism we
predict the masses of the exotic quarks in this model to be of the order of 
few TeV. 
 
\end{abstract}

\vskip 0.5in

The ${SU(3)_C\otimes SU(3)_L\otimes U(1)_X}$ extension \cite{pisano,frampton}
of the Standard model 
predicts interesting new physics at the TeV scale. In the minimal
version of the model one requires four Higgs multiplets in order
to generate experimentally acceptable mass spectra. The scale of
symmetry breaking, $M_Y$, is in the range of several hundred
GeV to a few TeV 
and is fixed by the value of $\sin^2({\theta}_W)$ which at this scale is
given by,
$ 4 \sin^2({\theta}_W) =  1/\left(1 +  g^2/(4{g^2_X})\right),\cite{joglekar} $
g and $g_X$
 are the $SU(3)_L$ and $U(1)_X$ coupling constants.
The scale $M_Y$ is 
fixed by evolving this coupling so that its value at the 
electroweak symmetry breaking scale agrees with the experimental result.

The model predicts
five new gauge particles: one neutral $Z'$, which is dominantly the
U(1)$_X$ gauge boson and four charged bileptons. 
It also predicts 
three new exotic quarks with charges $-4/3$, $-4/3$ and $5/3$ respectively for
the three generations. The mass of the neutral gauge boson $Z'$ is 
experimentally constrained to be above 1.7 TeV \cite{joglekar}.
This
also requires that the gauge coupling for the U(1)$_X$ interaction is 
relatively large. For example, if $M(Z') = 1.7$ TeV, then the U(1)$_X$ 
coupling $g_X$
is roughly 1.5 and increases 
with further rise in the mass of $Z'$. This raises the possibility 
that this coupling may in fact be strong enough to dynamically break
the gauge symmetry of this model without requiring the introduction
of fundamental Higgs particles. In the present paper we investigate
this possibility and determine its phenomenological consequences.    

The fermion representations \cite{frampton} consist of
the lepton triplets,
\[\left( \begin{array}{c}
e \\ \nu_e\\ e^c\\ \end{array}\right)_L \quad ,\quad
\left( \begin{array}{c} 
\nu \\ \nu_\mu\\ \mu^c\\ \end{array}\right)_L \quad ,\quad
\left( \begin{array}{c}
\tau \\ \nu_\tau\\ \tau^c\\ \end{array} \right)_L\quad ,
\]
which transform as $\underline 3^*$ under SU(3)$_L$ with the U(1)$_X$ hypercharge
equal to 0, and the quark triplets
\[\left( \begin{array}{c}
u \\ d\\ D\\ \end{array}\right)_L \quad ,\quad
\left( \begin{array}{c} 
c \\ s \\   S\\ \end{array}\right)_L \quad ,\quad
\left( \begin{array}{c}
b \\ t\\   T\\ \end{array} \right)_L\quad ,
\]
transforming as $\underline 3$, $\underline 3$ and $\underline 3^*$ respectively.
The U(1)$_X$ hypercharges of these three quark triplets are $2/3$, $2/3$ and
$-4/3$ respectively. The right handed quarks are singlets under SU(3) with
U(1)$_X$ hypercharges $Y(u_R)=Y(c_R)=Y(t_R) = -4/3$, $Y(d_R)=Y(s_R)=Y(b_R) = 2/3$,
$Y(D_R)=Y(S_R) = 8/3$ and $Y(T_R) = -10/3$.
Note that the third quark generation is treated asymmetrically from
the first two. 
The
third component of the quark generations corresponds to new exotic quarks. 
The current experimental lower limits on the 
exotic quarks are
200 GeV \cite{das}. The  
new charged gauge bosons $Y^+$
and $Y^{++}$ in this model are constrained to have masses above
 270 GeV by \cite{carlson}. Neutrino oscillations further constrain
these masses to be above 300 GeV \cite{mckay}. Muonium-antimuonium conversion data,
however, give the most stringent limit of 800 GeV \cite{willman}.  

One of the unique and very interesting feature of this model is that
it requires atleast three generations of fermions
for anomaly cancellation, hence providing
a justification for their existence.  
The model has been well studied  
in the literature and has been found to
lead to phenomenological predictions which are consistent
with current experimental data. It predicts interesting new physics at
the next generation of colliders such as lepton number violation due to the 
existence of 
bilepton gauge bosons.

The experimental lower bound of 1.5 on $g_X$ leads to the 
Landau pole in the TeV scale. 
The one loop evolution of the U(1)$_X$ coupling   
${\alpha}_X$  
leads to  the Landau pole at,
   $Q = \mu \exp\left(
{\frac {24 \pi^2} {g_X^2{\left[\sum Y^2\right]}}}\right)$. 
Here the summation over $Y^2$ 
of all the quarks is equal to $288/9$.
For $g_X(M_{Z'}) = 1.53$, ($M_{Z'}=1.7$ TeV) leads to Landau pole at 40 TeV,  
which indicates the strong nature of the interaction and that we may 
expect strong dynamical effects at 
TeV energies. 

The pattern of dynamical symmetry breaking can be obtained by analysing
the Schwinger-Dyson equation for quarks. We will include only the
U(1)$_X$ interaction for this purpose since this gives dominant  contribution. 
Furthermore we confine ourselves to the rainbow approximation. 
For a quark with left hypercharge $Y_L$ and right hypercharge $Y_R$ the 
corresponding equation for the self energy can be written as,
 
\begin{equation}
\Sigma(q) =   \int \frac {d^{4}k}{{(2\pi)}^4} 
\Gamma_\mu {\frac {i}{\slash{k} - {\Sigma(k)}}}\Gamma_\nu
{\frac {\left(-g^{\mu\nu}+(k-q)^\mu (k-q)^\nu/(k-q)^2\right)}
{(k-q)^2 - M_{Z^\prime}^2}} \ .
\end{equation}
The 
dressed propagator $S(q) = i/(\slash{q}A(q^2) - \Sigma(q))$  and 
$q$ and $(k-q)$ are the four momentum of the Fermion and the Gauge
particle. In the Schwinger-Dyson equation we have set the wave function
renormalization $A(q^2) = 1$. 
 Going beyond this approximation requires a much more extensive
calculation \cite{jain}, which is unnecessary for our purpose. 
 If there exists a solution to 
this equation such that
$\Sigma(q) \ne 0$, then chiral symmetry is dynamically broken. 
Assuming that such a solution is indeed obtained then the gauge boson
$Z'$ will become massive. Since we are interested in obtaining a nonperturbative
solution we have included a $Z'$  mass term in this equation. It 
remains to be verified later that the  $Z'$ mass  generated by this 
mechanism is equal to the mass assumed while solving this equation. 
The form of $\Gamma_\mu$ used in Eq. (1) is given by,
 $\Gamma_\mu = {g_V}{\gamma_\mu} + {g_A}{\gamma_\mu}{\gamma_5}$,
$ g_V 
= {\frac
 {g_X}{2}}{\frac {({Y_L} + {Y_R})}{2}}$, 
$g_A = {\frac {g_X}{2}}{\frac {(-{Y_L} + {Y_R})}{2}}$.
The final equation for the quark self energy $\Sigma$ is,

\begin{equation}
\Sigma(q) = -i{1\over 2}(g_X^2Y_LY_R)
\int \frac {d^{4}k}{(2\pi)^4} \frac
 {\Sigma(k)}{[k^2 - \Sigma{^2}(k)]
[(k-q)^2 - M_{Z^\prime}^2]}
\end{equation}

The relevant coupling factor is therefore $g_X^2Y_LY_R$. We find that 
the factor $Y_LY_R$ is equal to $40/9$ for the third generation exotic
quark $T$ and $16/9$ for the other two exotic quarks as well for the
top quark. For the remaining quarks this factor is considerably smaller
or negative being $4/9$ for down and strange quarks and $-8/9$ for 
up, charm and bottom quarks. Based on these hypercharges we will show 
that only the three exotic quarks and the top quark can be above the critical
value 
for dynamical symmetry breaking. This will lead to a mass pattern where
the exotic $T$ quark is very heavy and the other two exotic quarks
have masses comparable to top with the rest of the quarks remaining 
massless. The reason why only four of these quarks acquire dynamical masses is
because  the effective coupling for the top quark has to be very close
to the critical coupling necessary for dynamical symmetry breaking in 
order to maintain the large ratio $\wedge/m_t$, where $\wedge$ is the 
scale of $SU(3)_L \otimes U(1)_X$ breaking which is expected to be of the order
of a few TeV.
In order to assure that at least the top and the exotic quarks corresponding
to the first two generations acquire dynamical masses, we have to demand
that the effective coupling is slightly above the critical value. A certain
amount of fine tuning is required  in order to maintain the small
mass of the top quark. The fine tuning required is  not excessive since the mass
of top is only one order of magnitude smaller 
than the scale of $SU(3)_L \otimes U(1)_X$
breaking. The effective couplings of the remaining quarks are therefore 
necessarily below the critical value required for dynamical symmetry breaking.

In order to get an estimate of the dynamically generated fermion and 
gauge boson masses  
we numerically solve the gap equation in the ladder approximation,
imposing an ultraviolet cutoff $\Lambda_c$ on this equation. 
If the gap equation accepts a 
$M_T \ne 0$ solution, where $M_T$ is the $T$ quark mass, 
then the gauge symmetry is dynamically broken 
and the $Z'$, $Y^{\pm}$, $Y^{\pm\pm}$ gauge bosons become massive with
masses equal to $g_XF_\Pi(Y_R - Y_L)/4$, $gF_\Pi/2$ and $gF_\Pi/2$
respectively, 
where $F_\Pi$ is the pseudoscalar decay constant and 
we calculate it using the
Pagels-Stokar approximation \cite{pagels},
which should be sufficient for our purpose. More detailed analysis \cite{jain}
requires considerable more numerical effort and is expected
to give results within 20\% of this formula. 
A consistency requirement imposed on our 
solution is that the mass
of the exchanged particle $Z'$ has to be equal to the mass generated by
the dynamical mechanism. 
The solutions to the gap equation are therefore
iteratively improved by starting with a trial guess for the 
exchanged boson mass and
then comparing it with the predicted mass obtained using the decay constant. 
Solutions to this equation for different cutoff and coupling $g_X$
choice are shown in Table 1.

Next we consider the top quark and the first two generation 
exotic quarks (D,S).
The product of hypercharges in this case is $Y_RY_L=16/9$ which is
considerably smaller than the corresponding value for $T$. Since the
mass of $T$ quark is in the TeV regime and
mass of top $m_t=175$ GeV, the effective coupling for the case
of top has to very close to critical value and requires a certain
fine tuning. 
 However since the ratio of the these two scales is only about 0.1,
the fine tuning required is not severe.
Since the effective coupling of the exotic quarks belonging to
the first two generations is identical to that of the top quark,
we predict that the mass of these exotic quarks should
be of the same order to that of top. 

The
degeneracy between the mass of top and these two exotic quarks
is broken after we include the effect of $Z$ boson and
photon.
We include these contributions also in the rainbow approximation.
The contribution of Z as well as the photon to the exotic quark
mass is larger than to top. This is because the electric charge as well
as the effective Z coupling factor of
these quarks is larger in comparison to the top.
These coupling, although
relatively small compared to the strong U(1)$_X$ coupling, give a substantial
contribution to the top, $D$ and $S$ quark masses. 
This is because the effective U(1)$_X$ coupling of these quarks is close to
critical value and even the small 
effects of including the Z boson and photon makes a significant contribution. 

Once the $t\bar t$ condensate is formed 
we get a direct prediction for the mass of the electroweak
gauge bosons $W$ and $Z$, assuming that there is no fundamental
Higgs interaction present in the model. 
The situation here is very similar to the one studied in Ref. \cite{bhl} 
except that in our case we do not need to introduce any four fermi 
interaction by hand.
Unfortunately the prediction turns out to 
be about half the experimental masses. For the range of parameters given
in Table 1, we find that the mass of W boson ranges from 30 GeV to 40 GeV. 
 An analogous situation was found
in \cite{bhl} where the authors had to choose very large cutoffs 
in order that with a top quark mass of around 175 GeV, the experimental
value of electroweak gauge particles is obtained. The relevant scale
in our model cannot however be much larger than a few TeV and hence our
prediction for $m_W, m_Z$ is necessarily much smaller. 
This might indicate the need for introducing some fundamental Higgs
multiplets or further modifying the gauge structure of the model.
The final result will also depend to some extent on the precise
truncation scheme chosen for the integral equations. It may be better
to, for example, also include the scale dependence of the U(1)$_X$ coupling since
it varies rapidly in this region. However we have tested the sensitivity of 
our prediction by including this scale dependence as well as by using 
a truncation scheme of the coupled Schwinger-Dyson and Bethe-Salpeter 
equation proposed in Ref. \cite{jain} 
and find that it changes by less than 20 \%.  
The prediction of the decay constant and hence of the W boson mass
is thus expected to be quite reliable.

The lower prediction for the W boson mass in this model is  
expected since it arises in all models which propose to break 
electroweak symmetry
purely by top condensation \cite{bhl,cvetic,hill}. 
A solution to this problem has recently been 
proposed \cite{dobrescu,chivukula} which lowers the mass
of the top quark through
a seesaw mechanism.
This arises due to mixing of the top quark
with another particle which
is a weak SU(2) singlet. An analogous 
mechanism can be implemented
in our model by introduction of new quarks $\chi_L$ and
$\chi_R$ which transform as $(3,1,-4/3)$  under  
${SU(3)_C\otimes SU(3)_L\otimes U(1)_X}$. 
The model 
remains anomaly free after introduction of these fermions.
Since $\chi$ do not couple to $SU(3)_L$, we can add the
mass terms $m_{\chi\chi}\overline\chi_L\chi_R + m_{\chi t}\overline
\chi_L t_R + h.c. $ to the lagrangian without explicitly breaking the 
gauge symmetry.
Following Ref. \cite{dobrescu,chivukula}
the mass matrix of the top, $\chi$ quarks is given by  
$\left(\matrix{m & m\cr m_1 & m_2}\right)$, where $m$ is the mass of
top quark in the absence of the quark $\chi$ and the value of $m_2,m_1(m_2>m_1>m)$
is determined by gauge coupling and the explicit mass parameters
$m_{\chi\chi}$ and $m_{\chi t}$. 
After diagonalization
the top quark mass is given by $m_t=m(1-m_1/m_2) < m$
and the ratio of $m_t/F_\Pi$ can be reduced by adjusting the
explicit masses $m_{\chi\chi}$ and $m_{\chi t}$ in order
to obtain the experimental result for W and Z masses. 
The results shown in Table 1 include the contribution of
the quark $\chi$. The ratio $m_1/m_2$ ranges from 0.48 to 0.30
for the results shown in Table 1.
We therefore find that the present model naturally gives an
explanation observed mass spectrum. It provides us with a 
simple and well motivated model to display the top condensation
mechanism.

Current experiments
constrain the masses of exotic quarks to be higher than 200 GeV \cite{das}.
Our prediction for the mass of the lightest exotic quark
is a few TeV which can only be probed in the next generation of
colliders. There also exist stringent limits on the mass of the 
bilepton gauge bosons  $Y^{--}$ and $Y^-$. The most stringent bound
is obtained from muonium to antimuonium conversion of about 800 GeV
\cite{willman}.
We note that since we do not require that our model be perturbative
our upper limit on the mass of these particles is much higher, of
the order of a few TeV \cite{joglekar} and the model is not currently
ruled out. 
We also point out that this bound of 800 GeV is obtained with the
assumption that the CKM matrix for the coupling of leptons to SU(3) gauge
bosons $Y^{--}$ and $Y^{-}$ is essentially equal to identity. 
This assumption is quite reasonable since as shown in \cite{liu},
current experimental data already demands that the corresponding CKM matrix
be close to one. However there is one other possibility that is not
ruled out currently by experiments. This is the case of maximal
mixing in the first two generations with $V_{11}\approx V_{22}\approx 0$
and $V_{12}\approx 1$. Ref. \cite{liu} claim to rule this out also
at 95\% confidence level by using the 
upper limit on the mass of $M_Y$ to be 430 GeV. However, as long
as we do not impose the constraint that the theory remains  within the
perturbative regime, the limit on the $Y$ mass is much higher \cite{joglekar}.   
Hence we argue that this region of the paramater space is so far 
not thoroughly explored and some of the limits on this model 
claimed in the literature are not valid if the CKM matrix has this
form. This include, for example, the recent bound on $Y$ mass
obtained by using the muonium to anti-muonium conversion \cite{willman}. In the 
case of maximal mixing between first two generations the model predicts
zero conversion rate and hence in this case no limit can be imposed on $m_Y$
based on this process.

In conclusion, we have shown that 
the $SU(3)_L\otimes U(1)_X$ extension of the Standard
Model naturally displays the top condensation mechanism 
\cite{bhl,cvetic,hill,miransky,marciano,top}
for electroweak
symmetry breaking. This model has been well studied in the literature and
has several interesting features such as the requirement 
of three generations for anomaly cancellation. 
In the present paper we have shown that current 
experiments require the U(1) coupling in this model to be 
strong, which leads to dynamical breakdown
of $SU(3)_L\otimes U(1)_X$ to $SU(2)_L\otimes U(1)$ and finally to 
U(1)$_{EM}$ through top
condensation. Hence the model gives an explanation for the 
dominant features of the observed mass spectrum. We find that the dynamical
symmetry breakdown mechanism predicts the masses of the first two generation
exotic quarks to be of the order of a few TeV and should be observable
at LHC.   

\medskip
\noindent
{\large\bf Acknowledgements:} We thank Doug McKay for very useful comments.
This work was completed in part when PJ was visiting ICTP.
He thanks the ICTP staff for hospitality.

\begin{table}
\caption{
The dynamically generated gauge boson and heavy quark masses for different
U(1)$_X$ gauge coupling and cutoff. The dynamically generated top quark 
and $W$ boson mass is fixed to be 175 GeV  
and 80 GeV respectively.  The first two generation exotic quarks
$D,S$ are
degenerate. 
}

\begin{tabular}{|c|c|c|c|c|c|}
$M_{Z^\prime}$(TeV) & $M_Y$ (TeV) & $\Lambda_c$ (TeV) & $m_T$ (TeV)
& $g_X$ & $m_D$, $m_S$ (TeV)\\
\hline
2.13 & 0.26 & 42 & 5.7 & 2.576 & 0.78\\
\hline
3.07 & 0.37 & 90 & 8.2 & 2.540 & 1.0  \\
\hline
4.08 & 0.50 & 162 & 10.9 &  2.524 & 1.3\\
\hline
4.47 & 0.55 & 196 & 11.9 & 2.521 & 1.4\\
\hline
5.00 & 0.61 & 247 & 13.4 & 2.514 & 1.6\\
\hline
7.09 & 0.87 & 504 & 19.0 & 2.507 & 2.2 \\
\end{tabular}
\end{table}

\end{document}